\documentclass[12pt]{article}
\usepackage{latexsym}
\usepackage{amssymb}
\newcommand{\dintst}{\int \! \! \! \int d^{4}x \, d^{4}y}

\newcommand{\dintsof}{\int_0^{s_0} \! \! \int_0^{s'_0} \!ds\,\; ds'}

\begin{document}
\parindent 0pt
\thispagestyle{empty}
\mbox{ }
\rightline{UCT-TP-258/99}\\
\rightline{February 2000}\\
\vspace{2.0cm}
\begin{center}
\begin{Large}
{\bf Axial anomaly, vector  meson dominance and $\pi^{0} \rightarrow
\gamma \gamma$ at finite temperature\\}
\end{Large}
\vspace{.7cm}
{\bf  C. A. Dominguez $^{(a)}$, M. Loewe $^{(b)}$} \\
\vspace{.7cm}
$^{(a)}$Institute of Theoretical Physics and Astrophysics\\
University of Cape Town, Rondebosch 7700, South Africa\\
\vspace{.5cm}
$^{(b)}$Facultad de Fisica, Pontificia Universidad Catolica
de Chile\\
Casilla 306, Santiago 22, Chile\\
\vspace{.5cm}
\end{center}
\vspace{.5cm}
\begin{abstract}
\noindent
A thermal Finite Energy QCD Sum Rule is used to determine the
temperature behaviour of the $\omega \rho \pi$ strong coupling.
This coupling decreases with increasing $T$ and vanishes at the
critical temperature, a likely signal for quark deconfinement.
This is then used in the Vector Meson Dominance (VMD) expression for the
$\pi^0 \rightarrow \gamma \gamma$ amplitude, which is also found to
vanish at the critical temperature, as expected. This result supports the
validity of VMD at $T \neq 0$. However, if VMD would not hold at finite
temperature, then there is no prediction for the
$\pi^0 \rightarrow \gamma \gamma$ amplitude.
\end{abstract}
\newpage
The quest for the quark-gluon plasma (QGP) \cite{REV1} 
has prompted a great deal of interest in the
thermal behaviour of hadronic Green's functions, particularly those of
pions and vector mesons. Subsequent to early work proposing that the
imaginary part of all hadronic Green's functions should increase with
temperature \cite{PISA1}-\cite{GAMMAT}, results from a variety of models
have  confirmed this idea \cite{REV1}. However, no such general consensus
seems to exist regarding the temperature dependence of hadron masses, i.e
whether masses should increase, decrease, or remain constant, as
the temperature increases. For instance,
in the case of the rho-meson it has been argued \cite{PISA2}
that if Vector Meson Dominance (VMD) holds at finite temperature, then
$M_{\rho} (T_{c}) > M_\rho (0)$, where $T_c$ is the chiral-symmetry
restoration temperature, while if VMD breaks down there is no prediction.
At temperatures below this phase transition different models can
give opposite behaviours (see e.g. \cite{PISA1} and \cite{CAD1})
The validity (or not) of vector meson dominance
at finite temperature has a clear impact on the physics of the QGP, and
hence the importance of analyzing this issue from different viewpoints.
In this regard, it has been shown e.g.
that the electromagnetic pion form factor at $T \neq 0$, determined
directly from three-point function QCD sum rules \cite{3PFT},
i.e. without invoking VMD, is in good
agreement with the VMD expression with couplings determined independently
\cite{GRHOPI}, thus supporting the validity of VMD at finite temperature
(for other analyses see e.g. \cite{PISA2} and references therein).
Another window into this issue is offered by the decay
$\pi^{0} \rightarrow \gamma \gamma$. It is well known that at zero
temperature and in the chiral limit, the amplitude for this decay,
$F_{\pi \gamma \gamma}$,
is related to the Adler-Bell-Jackiw axial anomaly \cite{ABJ}, i.e.
%Eq.1
\begin{equation}
F_{\pi \gamma \gamma} f_\pi = \frac {1}{\pi} \; \alpha_{EM} \; ,
\end{equation}
where $f_\pi \simeq 93$ MeV is the pion decay constant.
It is also known that this anomaly is temperature independent \cite{ANT}.
However, as shown in \cite{PISA3}, this does not imply that the product
$F_{\pi \gamma \gamma} f_\pi$ is independent
of $T$, because the relation between the decay amplitude and
the anomaly no longer holds at finite temperature \cite{PISA3}. This is due
to the loss of Lorentz invariance. It is easy to see that if this were
not the case, then  $F_{\pi \gamma \gamma}(T)$ would diverge at the
critical temperature $T_c$, contrary to the expectation
$F_{\pi \gamma \gamma}(T_c) = 0$. Furthermore, assuming VMD, the
naive scenario would also likely imply the divergence of the strong
$\omega \rho \pi$ coupling at $T=T_c$, once again contrary to
expectations (we assume deconfinement and chiral-symmetry restoration
take place at about the same temperature). The precise temperature
dependence of $F_{\pi \gamma \gamma}(T)$ will certainly depend on
the dynamical model used in the calculation. In this paper we determine
the $\omega-\rho-\pi$ coupling using a thermal Finite Energy QCD Sum Rule
(FESR). To be more specific, the FESR fixes the ratio of this coupling
and the photon-vector meson couplings. 
Assuming VMD we can then determine $F_{\pi \gamma \gamma}(T)$.
We find that, in fact, $F_{\pi \gamma \gamma}(T_c)|_{VMD} =0$. Without being
a rigorous proof, this result lends
support to the validity of VMD at $T \neq 0$. The specific behaviour of
$F_{\pi \gamma \gamma}(T)$ depends on the behaviour of $f_\pi(T)$,
which is known \cite{BAR1}, as well as on $M_\rho (T)$ and
$M_\omega (T)$, which are model dependent. We discuss various possibilities
for the latter, and compare the result for $F_{\pi \gamma \gamma}(T)$
with other determinations.\\

In order to have the correct normalization, we begin with the
determination of $g_{\omega\rho\pi}$ at zero temperature . To this end
we consider the three-point function
%Eq.2
\begin{equation}
  \Pi_{\mu \nu} = i^2 \dintst  \; \langle 0 \left| T\left(
  J^{(\rho)}_\mu (x)
  J^{(\pi)}_5 (y) J^{(\omega)}_\nu (0)
       \right) \right|0 \rangle   \;
  e^{- i (px+qy)} \; ,
\end{equation}
where 
$J^{(\rho)}_\mu  = :\overline{u} \gamma_\mu  d:$,
$J^{(\pi)}_5  = (m_u+m_d) :\overline{d} i \gamma_5  u:$, and
$J^{(\omega)}_\nu  = \frac{1}{6} :(\overline{u} \gamma_\nu  u
+\overline{d} \gamma_\nu  d):$, $q = p' - p$,
and the following Lorentz decomposition will be used
%Eq.3
\begin{equation}
  \Pi_{\mu \nu} (p,p',q) = \epsilon_{\mu \nu \alpha \beta} \; p^{\alpha}
  p'^{\beta} \; \Pi (p^2,p'^2,q^2).
\end{equation}
In perturbative QCD, to leading order in the strong coupling, this
three-point function vanishes identically to leading order in the quark
masses, as it involves the $ {\it Tr} (\gamma_5 \gamma_\alpha
\gamma_\beta \gamma_\rho \gamma_\sigma \gamma_\tau) \equiv 0$. Hence,
the perturbative Green's function is of order $m_q^2$ and can be
safely neglected. The dimension-four gluon condensate term also does not
contribute on account of the same trace argument.
This leaves the leading non-perturbative contribution involving the
quark condensate 
%Eq.4
\begin{equation}
\Pi(p^2,p'^2,q^2)|_\mathrm{QCD} = \frac{1}{6} (m_u+m_d) \left(\langle
\overline{u}
u \rangle + \langle \overline{d} d \rangle \right)
\left( \frac{1}{p^2 p'^2}+ \frac{1}{p^2 q^2}+ \frac{1}{p'^2 q^2}
\right)\;,
\end{equation}
where $p^{2}$ and $p'^2$ lie in the deep euclidean region, and $q^2$ is
fixed and arbitrary. The SU(2) vacuum symmetry approximation
$\langle \overline{u} u \rangle \simeq \langle \overline{d} d \rangle
\equiv  \langle \overline{q} q \rangle$ will be adopted in the sequel.
The above result reduces to that of \cite{PAVER} after converting
to their kinematics.\\

Turning to the hadronic representation, and after inserting rho- and omega-
meson intermediate states, one obtains
%Eq.5
\begin{equation}
\Pi(p^2,p'^2,q^2) | _\mathrm{HAD}= 2 \; \frac{M^2_{\rho}}{f_\rho}
\frac{M^2_{\omega}}{f_\omega} \frac{f_\pi \mu_\pi^2}{q^2}
\frac{g_{\omega \rho \pi}}{(p^2-M^2_\rho)(p'^2-M^2_\omega)},
\end{equation}
where $f_\pi\simeq 93$ MeV, and the vector meson couplings
are defined as
%Eq.6
\begin{equation}
\langle 0 \left | J^{\rho}_\mu \right | \rho^+ \rangle
= \sqrt{2} \; \; \frac{M^2_\rho}{f_\rho} \;\epsilon_\mu \;,
\end{equation}
%Eq.7
\begin{equation}
\langle 0 \left | J^{\omega}_\mu \right | \omega \rangle
= \frac{M^2_\omega}{f_\omega} \; \epsilon_\mu \;,
\end{equation}
%Eq.8
\begin{equation}
\langle \rho (k_1,\epsilon_1) \pi (q) | \omega (k_2,\epsilon_2)
\rangle
= g_{\omega \rho \pi} \epsilon _{\mu \nu \alpha \beta} \;
\epsilon_1 ^ {\mu} \epsilon_2 ^ {\nu} k_1^{\alpha} k_2^{\beta}
\end{equation}

In Eq.(5), the pion propagator has been written in the chiral limit. This
approximation is consistent with having used massless internal
quark propagators in the QCD calculations. In fact, the term
$f_{\pi} \mu_\pi^2$ in Eq. (5) equals $(m_u+m_d) \langle \overline{q}
q \rangle /f_{\pi}$ on account of the Gell-Mann, Oakes and
Renner (GMOR) relation . \\

Next, using Cauchy's theorem, and assuming quark-hadron duality, the
lowest dimensional FESR for $g_{\omega \rho \pi}$ reads
%Eq.9
\begin{equation}
\dintsof \;Im \; \Pi(s,s',q^2)_\mathrm{HAD} =
\dintsof \;Im \; \Pi(s,s',q^2)_\mathrm{QCD}\; ,
\end{equation}
where $s=p^{2}$, $s'=p'^{2}$, and $s=s_{0}$ and $s'=s'_{0}$, are the usual
continuum  thresholds. 
From this FESR one then obtains the relation
%Eq.10
\begin{equation}
g_{\omega \rho \pi} = \frac{1}{6} \frac{f_\rho}{M^2_\rho}
\frac{f_\omega}{M^2_\omega} f_\pi \frac{\left [ - (m_u + m_d)
\langle \overline{q}q \rangle \right]}{f^2_\pi \mu^2_\pi} \left(s_0 + s'_0
\right)
\end{equation}

This result is clearly not a prediction for $g_{\omega \rho \pi}$,
as $s_{0}$ and $s'_{0}$ are a-priori unknown. However,
since the double dispersion in $p^2 = s$ and $p'\;^2 = s'$
refers to the vector meson legs of the three-point
function, with $M_\rho \simeq M_\omega$,
it is reasonable to set $s_0 = s'_0$. Furthermore, using the
experimental values
\cite{PDG}:  $f_\rho = 5.1 \pm 0.3$ and $f_\omega = 15.7 \pm 0.8$,
together with $s_0$ in the  typical range: $\sqrt{s_0} \simeq
1.2 - 1.5$ GeV, Eq. (10) then leads to  $g_{\omega \rho \pi} \simeq
11 - 16 \; \mbox{GeV}^{-1}$, in good agreement with the value extracted
from  $\omega \rightarrow 3 \pi$ decay  ($g_{\omega \rho \pi}  \simeq 16
\; \mbox{GeV}^{-1})$), or the one extracted from 
$\pi^0 \rightarrow \gamma \gamma$
decay using VMD ($g_{\omega \rho \pi} \simeq 11 \; \mbox{GeV}^{-1})$)
\cite{CADG}. This level of agreement suffices, as we are not
interested here in a precision determination of $g_{\omega \rho \pi}$
but rather in its thermal behaviour, i.e.
we shall concentrate on the ratio 
$g_{\omega \rho \pi}(T)/g_{\omega \rho \pi}(0)$.\\

The extension of the above analysis to finite temperature is
straightforward, i.e. all parameters entering Eq.(10) become, in principle,
temperature dependent.
It has been shown recently \cite{GMORT} that there are no temperature
corrections to the GMOR relation at leading order in the quark masses. To
next to leading order the corrections are of order $m_{q}^2 T^2$, numerically
very small except near the critical temperature for chiral-symmetry
restoration. The temperature dependence of $s_{0}$ was first obtained in
\cite{DL1}, and later improved in \cite{BAR2}. It turns out that for a wide
range of temperatures not too close to $T_{c}$, say $T < 0.8 \;T_{c}$, the
following scaling relation holds to a good approximation
%Eq.11
\begin{equation}
\frac{f_{\pi}^{2}(T)}{f_{\pi}^{2}(0)} \simeq
\frac{\langle \bar{q} q\rangle _{T}}{\langle \bar{q} q\rangle _{0}} \simeq
\frac{s_{0}(T)}{s_{0}(0)} \;.
\end{equation}
Hence, Eq. (10) can be recast as
%Eq.12
\begin{equation}
\frac{G(T)}{G(0)} \equiv \frac{g_{\omega\rho\pi}(T)/f_\rho (T) f_\omega (T)}
{g_{\omega\rho\pi}(0)/f_\rho(0) f_\omega(0)}=
\frac{f_\pi^3(T)}{f_\pi^3(0)}
\frac{1}{M_\rho^2(T)/M_\rho^2(0)}
\frac{1}{M_\omega^2(T)/M_\omega^2(0)} \; .
\end{equation}
The function $G(T)$ above is precisely the ratio appearing in the VMD
expression for the $\pi^0 \rightarrow \gamma \gamma$ decay amplitude
$F_{\pi\gamma\gamma}$ of Eq.(1), viz.
%Eq.13
\begin{equation}
F_{\pi\gamma\gamma}|_{VMD} = 8 \pi \alpha_{EM} \frac{g_{\omega\rho\pi}}
{f_\rho f_\omega} \; ,
\end{equation}
so that
%Eq.14
\begin{equation}
\frac{F_{\pi\gamma\gamma}(T)}{F_{\pi\gamma\gamma}(0)}|_{VMD}
= \frac{G(T)}{G(0)} \;.
\end{equation}
While the temperature dependence of $f_\pi$ is well known analytically
\cite{BAR1},
this is not the case for the vector meson masses. We discuss then the
behaviour of $G(T)$ according to various possibilities for the thermal
vector meson masses. (a) If $M_\rho(T) \simeq M_\rho(0)$ and
$M_\omega(T) \simeq M_\omega(0)$ then $G(T)$ vanishes as $f_\pi^3(T)$
as $T \rightarrow T_c$; (b) If $M_\rho(T) > M_\rho(0)$ and
$M_\omega(T) \simeq M_\omega(0)$ then $G(T)$ still vanishes as $f_\pi^3(T)$;
(c) If both $M_\rho(T)$ and $M_\omega(T)$ vanish at $T=T_c$ as
$f_\pi(T)$, then
$G(T)$ diverges as $1/f_\pi(T)$; the latter being a trivial property
of the bag model, where everything scales as $f_\pi(T)$.
Possibility (b) has been argued to be
a consequence of VMD \cite{PISA2}. It is then rewarding to see that
in this case $F_{\pi\gamma\gamma}(T)|_{VMD}$ vanishes at $T=T_c$, a behaviour
to be expected qualitatively on general grounds, and quantitatively
in specific field theory models \cite{PISA3}. At first sight it would appear
that possibility (c) contradicts the expectation that 
$F_{\pi\gamma\gamma}(T_c) = 0$.
However, this is not necessarily the case because such a behaviour for
the vector meson masses implies that VMD is no longer valid at finite
temperature \cite{PISA2}, in which case Eq.(14) does not have to follow.\\

Regarding the thermal behaviour of $g_{\omega\rho\pi}$ itself, 
in addition to its dependence on the thermal vector meson masses,
it also depends on how do $f_\rho$ and $f_\omega$ change with temperature.
Intuitively, one
would expect a decoupling of currents from hadrons at the critical
temperature. This is confirmed e.g. by the behaviour of the current-nucleon
coupling \cite{MNT}, and of $f_\rho$ \cite{LEE} (in chiral models
$f_\omega$ is temperature independent at leading order because the
omega meson does not couple to two pions). The coupling
$g_{\omega\rho\pi}(T)$ would then vanish as $f_\pi^3(T)$ if
$f_\rho(T) \simeq f_\rho(0)$ and $f_\omega(T) \simeq f_\omega(0)$,
or faster than $f_\pi^3(T)$ if $f_\rho(T_c)=f_\omega(T_c)=0$, for both
possibilities (a) and (b) above. In case (c) $g_{\omega\rho\pi}(T)$
would still vanish as $f_\pi(T)$ because  $f_\rho(T)$ and $f_\omega(T)$
would scale as the vector meson masses.\\

In summary, the thermal FESR used here to obtain the function $G(T)$
in Eq.(12) leads to $G(T) \rightarrow 0$ as $T \rightarrow T_c$,
and assuming VMD it leads also to the vanishing of the
$\pi^0 \rightarrow \gamma \gamma$ amplitude $F_{\pi \gamma \gamma}$,
provided the vector meson masses $M_\rho$ and $M_\omega$ do not vanish
simultaneously at $T=T_c$. If the latter would be the case, then VMD
does not hold at finite temperature \cite{PISA2}, so that Eq.(14) does
not necessarily follow. We view the above result as supporting evidence for
the validity of VMD at $T \neq 0$. Finally, the strong coupling
$g_{\rho \omega \pi}$ vanishes at the critical temperature, regardless
of the thermal behaviour of the vector meson masses. This may be interpreted
as analytical evidence for quark deconfinement.

{\bf Acknowledgements}\\
This work has been supported in part by the NRF (South Africa), and by
FONDECYT (Chile) under contracts Nos. 1980577 and 7980011. \\

\end{document}